\documentstyle[twoside,fleqn,espcrc2,epsf,rotate]{article}

\newcommand{\AmS}{{\protect\the\textfont2
  A\kern-.1667em\lower.5ex\hbox{M}\kern-.125emS}}

\title{Computers for Lattice Field Theories}

\author{Y. Iwasaki \address{Institute of Physics, University of Tsukuba,
        Ibaraki 305, Japan}
       }       
\begin{document}

\begin{abstract}
Parallel computers dedicated to
lattice field theories are reviewed with emphasis on the three recent
projects, the Teraflops project in the US, the CP-PACS project in Japan and 
the 0.5-Teraflops project in the US.
Some new commercial parallel computers are also discussed.
Recent development of semiconductor technologies is briefly
surveyed in relation to possible approaches toward Teraflops computers.
\end{abstract}

\maketitle

\section{Introduction}
Numerical studies of lattice field theories have developed significantly
in parallel with the development of computers during the past decade.
Of particular importance in this regard has
been the construction of dedicated QCD computers (see Table 1 and for earlier
reviews see Ref.\cite{review}) and 
the move of commercial vendors toward parallel
computers in recent years.  Due to these 
developments we now have access to parallel computers which are capable of
5--10 Gflops of sustained speed.  

However, a fully convincing numerical
solution of many of lattice field theory problems, in particular those of
lattice QCD, requires much more speed.  
In fact typical number of floating point operations required in these 
problems, such as full QCD hadron mass spectrum calculations,
often exceeds
$10^{18}$, which translates to 115 days of computing time with the sustained
speed of 100 Gflops.   
Under this circumstance 
we really need computers with a sustained speed exceeding 100 Gflops.

\begin{table}[t]
\setlength{\tabcolsep}{0.5pc}
\newlength{\digitwidth} \settowidth{\digitwidth}{\rm 0}
\catcode`?=\active \def?{\kern\digitwidth}
\caption{List of dedicated QCD computers}
\label{tab:machines}
\begin{tabular*}{7.4cm}{@{}r@{\extracolsep{\fill}}rr}
\hline

Project & \, \, Peak speed & year\\
   &  Gflops& \\
\hline
\hline
 Columbia 16& 0.25&1985\\
  64     &   1.0        & 87   \\
  256     &  16         & 89   \\
\hline
APE 4       &  0.25         &  86  \\
16       &  1.0         & 88   \\
\hline
QCDPAX       & 14          & 90   \\
\hline
GF11       &  11         & 91   \\
\hline
ACPMAPS       & 5          & 91   \\
\hline
\hline
APE100 &$ 6(\rightarrow 100)$&$92(\rightarrow 94)$\\
\hline
ACPMAPS &50&93\\
upgraded& & \\
\hline
\hline
Teraflops&1,600&96\\
\hline
CP-PACS&$ \geq300$&96\\
\hline
0.5Teraflops&800&95\\
\hline
\hline
APE1000&$\sim 1000$& \\
\hline
\end{tabular*}
\end{table}

In this talk I review the present status of effort toward 
construction of dedicated parallel computers with the peak speed of
100--1000 Gflops.  Of the six projects in this category (see Table 1),
APE100\cite{ape100} is near completion and 
ACPMAPS upgraded\cite{acpmaps_upgraded} is running now.   
Because they have already been reviewed
previously\cite{review}, 
we shall only describe their most recent status.   The
three recent projects, the Teraflops project\cite{teraflops_old,negele} 
in the United States, 
the CP-PACS project\cite{oyanagi} in Japan and
the 0.5-Teraflops\cite{christ} project in the United States,
are at a varying stage of development.
I shall describe them in detail.  Finally the APE1000\cite{rapuano} is the future
plan of the APE  Collaboration,  of which details are not yet available.

\begin{figure}[t]
\begin{center}
\leavevmode
\epsfxsize=6.3cm \epsfbox{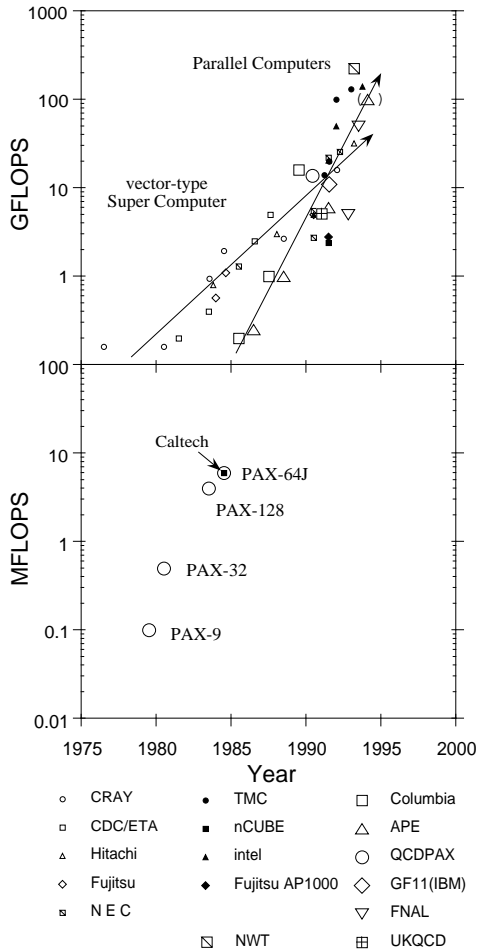}
\vspace{-0.8cm}
\end{center}
\caption{Progress of theoretical peak speed.}
\label{fig:ComputerSpeed}
\end{figure}
A key ingredient in the fast progress of parallel computers in recent years is
the development in semiconductor technologies.  Understanding this aspect is
important when one considers possible approaches toward a Teraflops of
speed.  I shall therefore start this review with a brief reminder of the
development of vector and parallel computers and the technological reasons why
recently parallel computers have exceeded vector computers in the computing
capability (Sec. 2).  The status of APE100 and ACPMAPS upgraded 
are summarized in Sec. 3.
The US Teraflops, CP-PACS and 0.5-Teraflops projects are described in Sec. 4. 
Powerful parallel computers are also available from commercial vendors.  In
Sec. 5 I shall discuss two new computers, the Fujitsu VPP500 and CRAY T3D. 
After these reviews I discuss several architectural issues for computers 
toward Teraflops in Sec. 6.
A brief conclusion is given in Sec. 7.

\section{Recent development of computers and semiconductor technology}

In the upper part of Fig.~\ref{fig:ComputerSpeed} we show the progress of
peak speed of vector and parallel 
computers over the years. Small symbols correspond to the first shipping
date of computers made by commercial vendors, with open ones for vector  and
filled ones for parallel type.  Parallel computers dedicated to lattice QCD are
plotted by large symbols.  We clearly observe that the rate of progress for
parallel computers is roughly double that of vector computers and that a
crossover in peak speed has taken place from vector to parallel computers around
1991.

The ``linear fit'' drawn in Fig.~\ref{fig:ComputerSpeed} for parallel
computers can be extrapolated to the period prior to 1985.  QCDPAX is the fifth
generation computer in the PAX series\cite{hoshino}
and there are four earlier computers
starting in 1978.  In the lower part of Fig.~\ref{fig:ComputerSpeed} the peak speed of these computers
are plotted in units of Mflops
together with that of the Caltech computer described, for example,
by Norman Christ
at Fermilab in 1988\cite{review}.  It is amusing to observe that the rapid increase of speed of
parallel computers has been continuing for over a decade since the early days.

\begin{figure}[t]
\epsfxsize=7.5cm \epsfbox{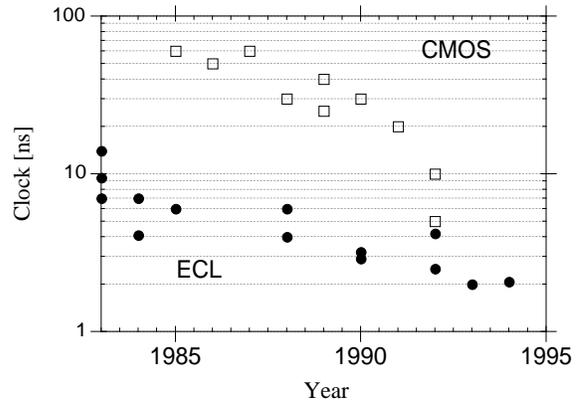}
\vspace{-0.5cm}
\caption{Machine clock of ECL and CMOS semiconductors.}
\label{fig:Clock}
\end{figure}
It is important to note that the first three PAX computers are limited to 8
bit arithmetic and the fourth one to 16 bit.  We also recall that the first
Columbia computer  used 22 bit arithmetic.  Thus not only the peak speed but also the
precision of floating point numbers has increased significantly for parallel
computers. Now the 64 bit arithmetic is becoming standard.

To see more closely why the crossover happened,
let us look at the development of technology of semiconductors.
In Fig.~\ref{fig:Clock} we show how machine clocks become faster in the case of
ECL which is utilized in vector-type supercomputers as well as in the  
the case of CMOS which is used in personal computers and workstations.
As we can see, the speed of CMOS is about 10-fold less
than ECL. However, the power consumption and the heat output are much
lower than those of ECL. Furthermore the speed of CMOS itself has become 
comparable to that of ECL of the late 1980's.

\begin{figure}[t]
\epsfxsize=7.5cm \epsfbox{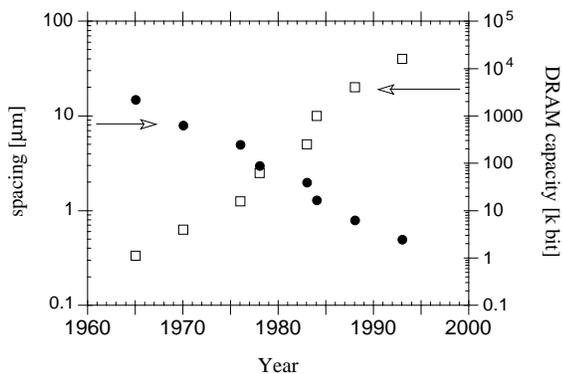}
\vspace{-0.5cm}
\caption{Development of minimum spacing of LSI and capacity of DRAM.}
\label{fig:Dram}
\end{figure}
\newbox\B
\epsfysize=7.0cm
\setbox\B=\vbox{\epsfbox[96 64 500 673]{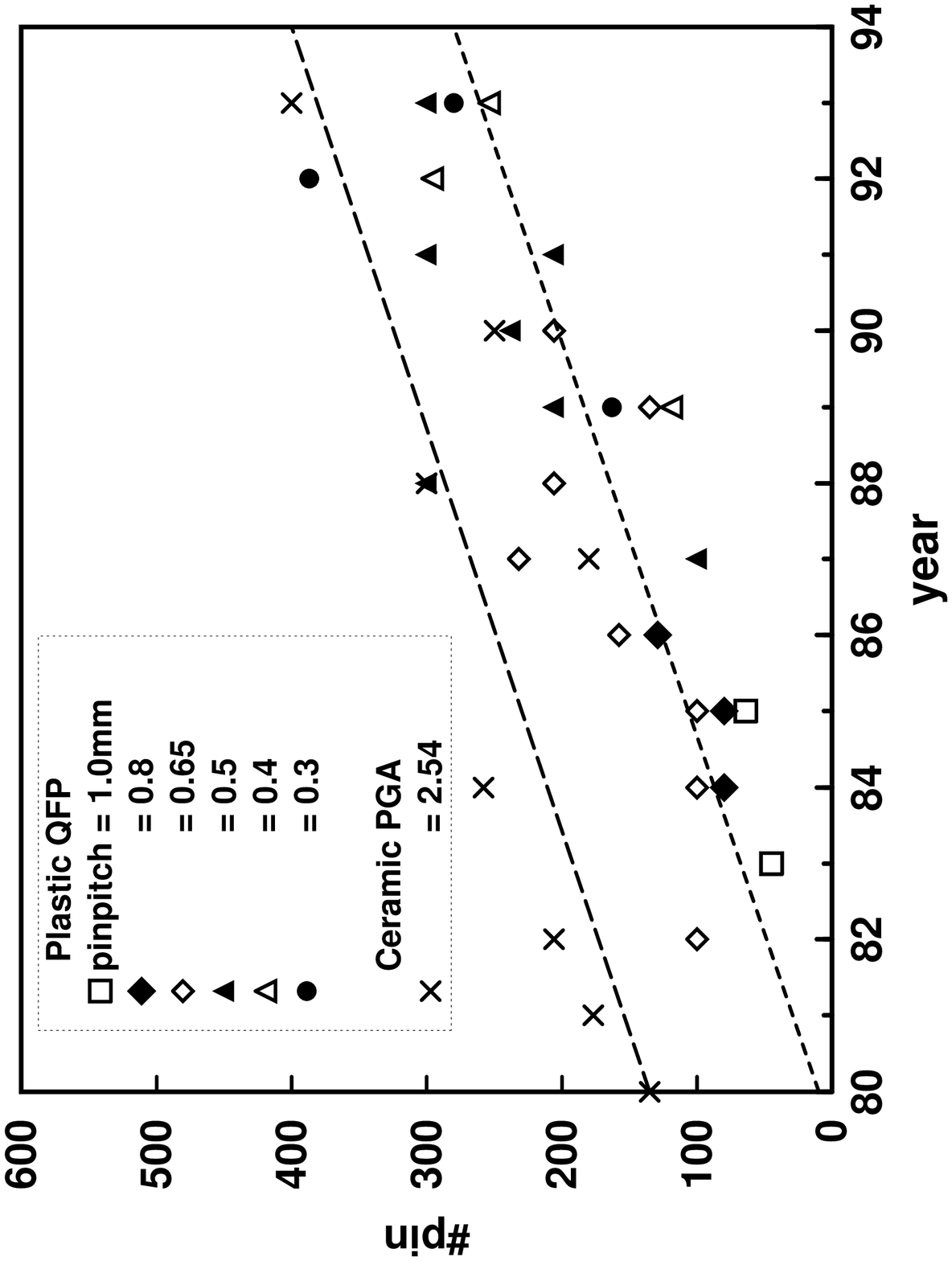}}
\begin{figure}[t]
\begin{center}
\leavevmode
        \makeatletter
        \@rotr\B
        \makeatother
\vspace{-0.5cm}
\end{center}
\caption{Evolution of the number of pins. (From ``Nikkei Electronics''
August 2, 1993.)}
\label{fig:Evol}
\end{figure}

\begin{table*}[t]
\setlength{\tabcolsep}{0.5pc}
\caption{Characteristics of dedicated QCD computers I}
\label{tab:QCD1}
\begin{tabular*}{\textwidth}{@{}l@{\extracolsep{\fill}}rrrrr}
\hline
Project & Columbia & APE & QCDPAX & GF11 & ACPMAPS  \\
\hline
peak       &           &           &          &          &   \\
speed & 16 Gflops & 1 Gflops & 14 Gflops & 11 Gflops & 5 Gflops \\
\hline
processors & 256  & 16  & 480 & 566  & 256     \\
\hline
network & 2d torus & linear & 2d torus & Memphis & crossbar and \\
        &        & array  &        & switch  & hypercube+     \\
\hline
arichi- &      &      &      &      &        \\
tecture & MIMD & SIMD & MIMD & SIMD & MIMD \\
\hline
CPU         & 80286 & --- & 68020 & --- & Weitek  \\
FPU         & 80287 & Weitek1032$\times$4 & LSI Logic 
            & Weitek1032$\times$2 
            & XL8032  \\
            & Weitek3364$\times$2 & Weitek1033$\times$4 & L64133 
            & Weitek1033$\times$2 
            & chip set        \\
\hline
SRAM        & 2MB & --- & 2MB & 64KB & 2MB    \\
DRAM        & 8MB & 16MB & 4MB & 2MB & 10MB \\
\hline
speed/ & & & & & \\
processor & 64Mflops & 64Mflops & 32Mflops & 20Mflops & 20Mflops\\
\hline
host    & VAX11/780 & $\mu$VAX & Sun 3/260 & 3090 & $\mu$VAX \\
\hline
\end{tabular*}
\end{table*}

The machine cycle of one nano-second is a kind of limit 
to reach. This is understandable because 
one nano-second is the time in which light travels 30cm.
In this time interval one has to load data from memory
to a floating point operation unit, make a calculation 
and store results to the memory.
Even in the ideal case of pipelined operations, one nano-second corresponds 
only to
one Gflops. Usually a vector computer has a multiple operation units
which consists of, for example, 8 floating point operation units (FPUs).
Because of this, the theoretical peak speed becomes 8 Gflops.
Further it has multiple sets of this kind of multiple FPUs; 
in the case of 4 sets the peak speed becomes 32 Gflops.
This is the way how a vector computer gets the peak speed of order of 10
Gflops.
That is, recent vector computers are already parallel computers.
However, it is rather difficult to proceed further in this approach
because of the power consumption and the heat output.

On the other hand, the development of CMOS semiconductor technology,
with its small-size, high speed and low power consumption,
has made it possible to construct
a massively parallel computer which is composed of order of 1,000 nodes with
the peak speed which exceeds that of vector-type supercomputers.
This is the reason why the crossover occurred.

The speedup of CMOS has become possible due to the development of 
LSI technology. 
Figure~\ref{fig:Dram} shows the development
in terms of the minimum feature size or minimum spacing. 
Now the spacing  has been reduced to 0.5 micron. 
This development has also lead to a substantial increase of 
DRAM bit capacity which has recently reached the level of 16Mbit.
The speed of transistors has also increased with the decrease of
minimum spacing because electrons can move through the minimum spacing
in a shorter time.
This is the reason why the machine clock has become faster.

The packaging technique has also developed: Figure~\ref{fig:Evol}
shows the development of the number of
pins of LSI.

Due to these development, it is now not a 
dream to construct a 1Tflops computer with 64 bit arithmetic
with reasonable size and reasonable power consumption.

\section{Past and present of dedicated computers}
The computers of the first group in Table~\ref{tab:machines},
the three computers of Columbia\cite{columbia}, two versions of APE\cite{ape}, 
QCDPAX\cite{qcdpax}, GF11\cite{gf11} and ACPMAPS\cite{acpmaps},
were constructed some years ago and have been producing physics results.
The characteristics of these computers are given in Table~\ref{tab:QCD1}.
These computers are already familiar to lattice community.
Therefore 
I refer to earlier 
reviews~\cite{review} for details and just emphasize that a number of
interesting physics results have been produced. 
This fact shows that there is really benefit in constructing dedicated 
computers.

The computers of the second group in Table~\ref{tab:machines}, 
the 6 Gflops version of APE100 and ACPMAPS upgraded,
have been recently completed. Both are now producing physics results,
some of which have been reported at this conference. 
I list their characteristics in Table~\ref{tab:QCD2}.

\subsection{APE100}
The architecture of APE100\cite{ape100} is a combination of SIMD and MIMD.
The full machine consists of 2048 nodes 
with a peak speed of 100 Gflops. The network is 
a 3-dimensional torus. 
Each node has a custom-designed floating point chip called MAD.
The chip contains a 32-bit adder and a multiplier with a 128-word register file.
The memory size is 4Mbytes/node with 80 ns access time
1M $\times 4$ DRAM. The bandwidth between MAD and the memory is 50 Mbytes/sec,
which corresponds to one word/4 floating point operations.
One board consists of $2 \times 2 \times 2$ = 8 nodes with a commuter for
data transfer.
The communication rates on-node and inter-node are 50 Mbytes/sec and
12.5 Mbytes/sec, respectively.
Each board has a controller which takes care of program 
flow control, address generation and memory control.

The 6 Gflops version of APE 100, which is called TUBE, is running and
producing physics results.
A TUBE is composed of 128 nodes making a $32 \times 2 \times 2$ torus
with periodic boundary conditions.
The naming originates from its topological shape.
The memory size is 512 Mbytes.
Four TUBEs have been completed.

The sustained speed of a TUBE
for the link update is about 1.5 microsecond/link
with the Metropolis algorithm with 5 hits. The time for multiplication of the Wilson
operator is 0.8 microsecond per site. These rates roughly
correspond to 2.5 Gflops
to 3 Gflops, which represents 40-50\% of the peak speed. These figures show
good efficiency.

The physics subjects being studied on TUBE are hadron spectrum and
heavy quark physics, the results of which have been reported at this conference.

A Tower which consists of 4 TUBEs with a peak speed of 25 Gflops
is being assembled now and should be working in the late fall of 1993.
The full machine  which is composed of 4 Towers with a peak speed of 100 Gflops
is expected to be completed by the first quarter of 1994.

\begin{table}[t]
\setlength{\tabcolsep}{0.5pc}
\caption{Characteristics of dedicated QCD computers II}
\label{tab:QCD2}
\begin{tabular*}{7.4cm}{@{}l@{\extracolsep{\fill}}rr}
\hline
Project & APE100 & ACPMAPS\\

        &      &\\
\hline
processors & 2048  & 612 \\
\hline
arichi- & SIMD     &    \\
tecture & MIMD & MIMD\\
\hline
CPU         & MAD & i860\\
            & (custom) & \\
\hline
memory      & 4MB & 32MB \\
\hline
speed/ &50 &80 \\
processor & Mflops & Mflops\\
\hline
network & 3d torus & crossbar\\
        &        & hypercube+\\
\hline
host    & SUN WS  & SGI \\
\hline
peak       &           &  \\
speed & 100 Gflops & 50Gflops\\
\hline
arithmetic        & 32 bit    & 32 (64) bit\\
\hline
\end{tabular*}
\end{table}

\subsection{ACPMAPS Upgraded}
This is an upgrade of the ACPMAPS replacing the processor boards
without changing the communication backbone\cite{acpmaps_upgraded}.
The ACPMAPS is a MIMD machine with distributed memory.
On each node there are two Intel i860 microprocessors with a peak speed of
80 Mflops. The memory size is 32 Mbytes of DRAM for each node.
The full machine consists of 612 i860 with a peak speed of 50 Gflops and
has 20 Gbytes of memory.

The network has a cluster structure:
one crate consists of 16 boards with
a 16-way crossbar. 
A board can be either a
processor node or a Bus Switch Interface board.
The 16-way crossbars are connected in a complicated way which makes
a hyper-cube and other extra connections.
The throughput between nodes is 20 Mbytes/sec.

ACPMAPS has a strong distributed I/O system:
there are 32 Exabyte tape drives and 20 Gbytes of disk space.
This mass I/O subsystem is one of characteristics of ACPMAPS.

The software package CANOPY which was well described several 
times\cite{acpmaps,acpmaps_upgraded}
is very powerful to distribute physical variables to nodes
without knowing the details of the hardware.

The ACPMAPS is running and doing calculations of the quenched hadron
spectrum and heavy quark physics, the results of which have been 
reported at this conference.

The sustained speed measured on a $32^3 \times 48$ lattice are as follows.
One link update time by a heat-bath method is 0.64 micro-second per link.
One cycle of conjugate gradient inversion of the Wilson operator 
by red-black method 
takes about 0.64
micro-second per site. The L inversion together with the U back-inversion
in the ILUMR method takes 2.23 micro-second
per site. These figures for the
sustained speeds are about 10-20\% of the peak speed.
Therefore efficiency is not so good compared to TUBE. 
However, there are several good characteristics. First, it supports
both 64 and 32 bit arithmetic operations. The network is very flexible and
the distributed I/O system is convenient for users.

\begin{table*}[t]
\setlength{\tabcolsep}{0.5pc}
\caption{Characteristics of dedicated QCD computers III}
\label{tab:QCD3}
\begin{tabular*}{\textwidth}{@{}l@{\extracolsep{\fill}}rrr}
\hline
Project &  Teraflops&CP-PACS &0.5Tflops  \\
        &     &  & \\
\hline
processors & 8K & 1--1.5K & 16K\\
\hline
arichi- &  &     &      \\
tecture &  &MIMD & MIMD\\
\hline
CPU         &  & enhanced & DSP\\
            &  & PA-RISC & TI        \\
\hline
memory      & 32MB & 64MB & 2MB\\
\hline
speed/ &200--300 &200--300  &50\\
processor & Mflops & Mflops & Mflops\\
\hline
network &  & hypercrossbar & 4d torus\\
        &  &        &    \\
\hline
host    &   & main frame  &SUN WS \\
\hline
peak      &         &          &        \\
speed & $\ge1.6$Tflops & $\ge$300Gflops & 0.8Tflops\\
\hline
arithmetic        &64bit  & 64 bit  & 32 bit  \\
\hline
\end{tabular*}
\end{table*}

\section{Project under way and proposed}
The three projects of the third group in Table~\ref{tab:machines}, 
the Teraflops project, the CP-PACS project and 
the 0.5-Teraflops project 
are well under way. 
The basic design targets are listed
in Table~\ref{tab:QCD3}.
\subsection{Teraflops project}
The Teraflops project\cite{teraflops_old} has changed significantly since last 
year.
The new plan (Multidisciplinary Teraflops Project)\cite{negele} 
utilizes Thinking Machine's next generation platform
instead of CM5 as originally planned. A floating point processing unit(FPU)
called an arithmetic accelerator is to be constructed with a peak speed
in the range of 200 -- 300 Mflops. One node consists of 16 such FPUs
plus one general processor, with a peak speed of more than 3.2 Gflops and
512 Mbytes of memory. 

The full machine consists of 512 nodes with a peak speed of at least
1.6 Tflops with 64 bit arithmetic.
The sustained speed is expected to be more than 1 Tflops.
A preliminary estimate for the cost of the full machine is
\$20 -- 25M. This project is the collaboration of the QCD Teraflops 
Collaboration\cite{teraflops_mem}, MIT Laboratory for computer science, Lincoln Laboratory and TMC.
Funding for the project began in the fall of 1992 with start-up funds 
provided by MIT. The proposal for the whole project will be submitted to 
NSF, DOE and ARPA this fall. The tentative schedule is to build a 
prototype node in 1994, a prototype system in 1995 and have the full system in operation in 1996.

\subsection{CP-PACS project}
We started the 
CP-PACS (Computational Physics by Parallel Array Computer Systems)
project last year\cite{oyanagi}. The CP-PACS collaboration currently consists
of 22 members\cite{cppacs-mem}, a half of them physicists and the other half 
computer scientists.

The architecture is MIMD with a 3-dimensional hyper crossbar which will be 
explained later. The target of the peak speed is currently at least 300 Gflops
with 64 bit arithmetic. We are making a proposal for additional funds
to increase this peak speed.
The memory size is planned to be more than 48 Gbytes.

The processor is based on a Hewlett-Packard PA-RISC processor.
This is a
super-scalar processor which can perform two operations concurrently. 
We enhance the processor to support efficient vector calculations. 
The peak speed of one processor
is 200 -- 300 Mflops. 
The enhancement will be described in detail
later. For memory we use synchronous DRAM, pipelined by multi-interleaving
banks and a storage controller. The memory bandwidth is one word per 
one machine cycle.

\begin{figure}[t]
\epsfxsize=7.5cm \epsfbox{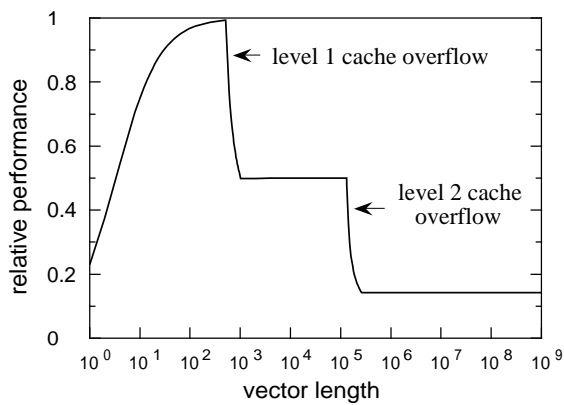}
\vspace{-0.5cm}
\caption{Performance of a  RISC processor in a large scale scientific calculation.}
\label{fig:Performance}
\end{figure}
Now let me explain the vector enhancement of the processor.
As is well-known, high performance of usual
RISC processors like those of Intel, IBM,
HP and DEC heavily depends on the existence of cache. 
However, when the data size exceeds the cache size, effectiveness of cache 
decreases.
Figure~\ref{fig:Performance} shows a typical example of the performance of a 
RISC processor.
When the data size exceeds the size of the on-chip level-1 cache, 
it drops down by about
50\%. 
Furthermore when it exceeds the size of the level-2 cache, 
the performance is of order 15\% of the theoretical peak speed. This feature is very common to
cache-based RISC processors.

To overcome this difficulty, our strategy is to increase the number
of floating-point registers without serious changes in the instruction set
architecture. This means upward compatibility. However, this is not 
straightforward
because the register fields for instructions are limited; 
the number of registers is usually
limited to 32. To resolve this problem we introduce slide windows
as well as preload and poststore instructions\cite{nakamura}. 
We also pipeline the memory.
Because of these features we are able to hide
long memory access latency and perform vector calculations efficiently.

\begin{figure}[t]
\epsfxsize=7.5cm \epsfbox{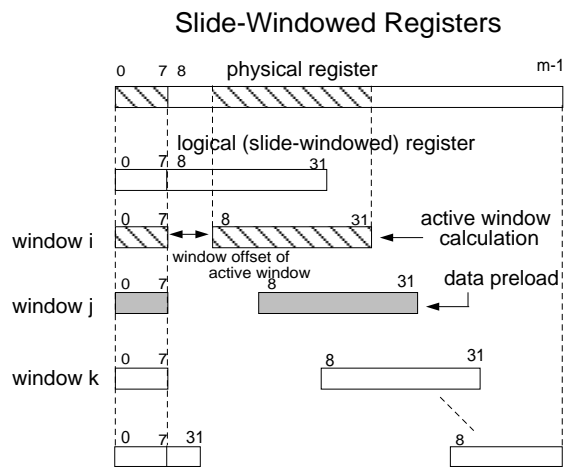}
\vspace{-0.5cm}
\caption{Schematic graph of slide-windowed registers.}
\label{fig:Slide-Window}
\end{figure}

Figure~\ref{fig:Slide-Window}
is a schematic illustration of how slide windowed registers work.
Arithmetic instructions  
use the registers in the active window which has 32 registers.
The preload instruction can load data into registers of the next (or next-to-next)
window and 
the poststore instruction stores data from registers of the previous
window. The pitch for the window slide can be chosen by software.
Due to the preload and poststore instructions we can use all of m ($m > 32$)
physical registers.

\begin{figure}[t]
\epsfxsize=7.5cm \epsfbox{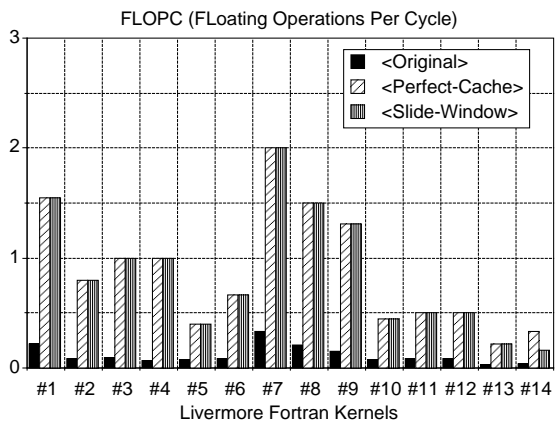}
\caption{Comparison of performance with and without slide windows for 
Livermore loops.}
\label{fig:Livermore}
\end{figure}
Figure~\ref{fig:Livermore} is a comparison of the performance with 
and without slide windows 
for Livermore Fortran Kernels : $<$Original$>$ means performance without slide
windows, and $<$Perfect-Cache$>$ represents a hypothetical case for comparison
where the cache size is
infinite and the data are all in cache.
In the case of $<$Slide-Window$>$,
the number of slide-windowed floating-point registers is assumed to be 64.
Except for \#14 of Livermore Fortran Kernels, 
the performance with slide windows is almost equal to that
of the perfect cache case and it is about 6 times higher than the original one.

\newbox\B
\epsfysize=7.0cm
\setbox\B=\vbox{\epsfbox[96 64 500 673]{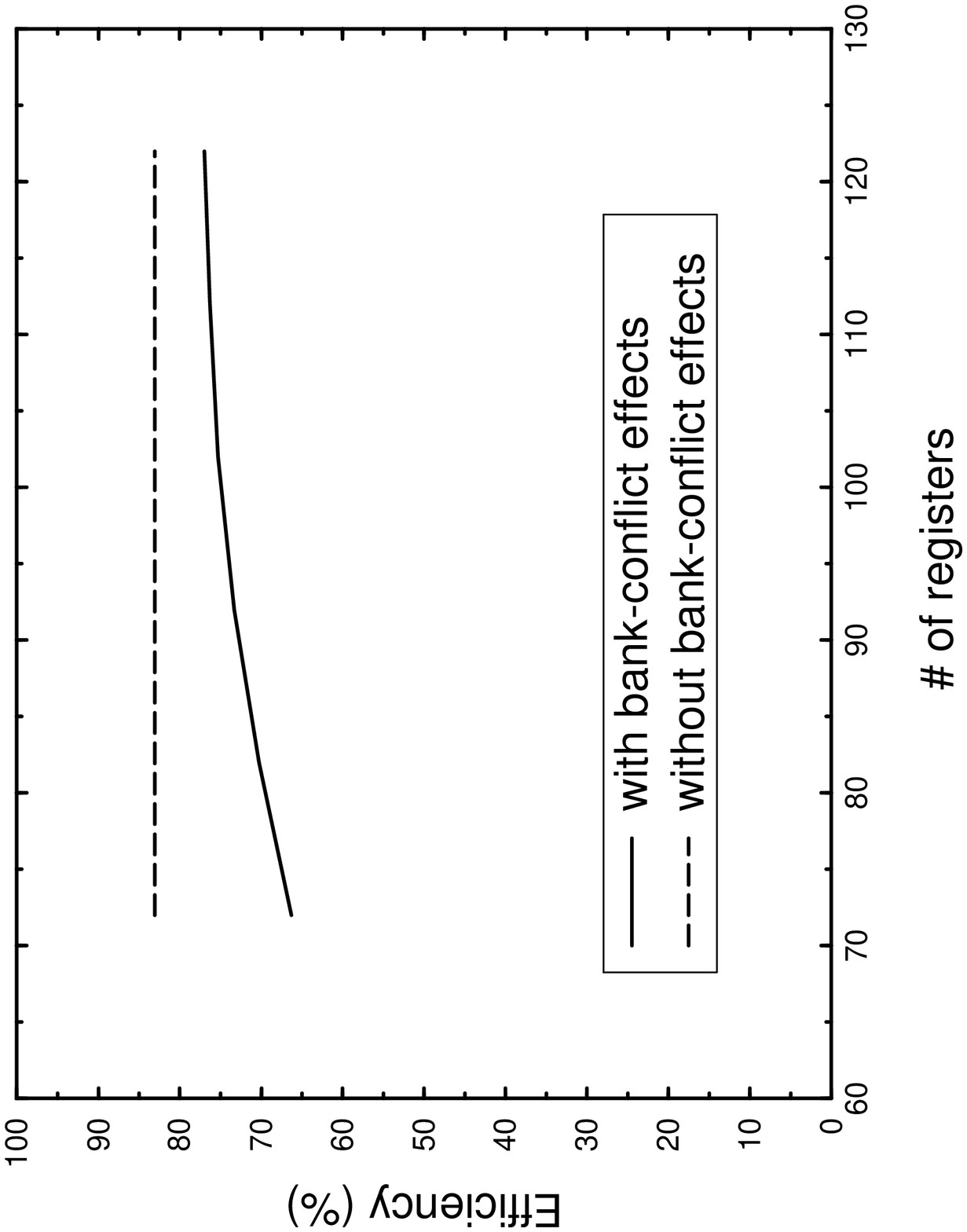}}
\begin{figure}[bt]
\begin{center}
\leavevmode
        \makeatletter
        \@rotr\B
        \makeatother
\end{center}
\caption{Performance for multiplication of Wilson matrix.}
\label{fig:Mult}
\end{figure}

Figure~\ref{fig:Mult}
shows the efficiency of performance for the case of multiplication of 
the Wilson matrix. The dashed line corresponds to efficiency in the 
case of the code optimized by hand without considering memory bank-conflicts.
The solid line is the result of a simulation for the realistic case where
the effect of memory bank conflict and the buffer size effect are taken into account.
This shows that if the number of registers is larger than 100
the efficiency is more than 75\%.
We will develop a compiler 
for the enhanced RISC processor,
which will produce optimized codes for the slide-window architecture.

\begin{figure}[t]
\epsfxsize=7.5cm \epsfbox{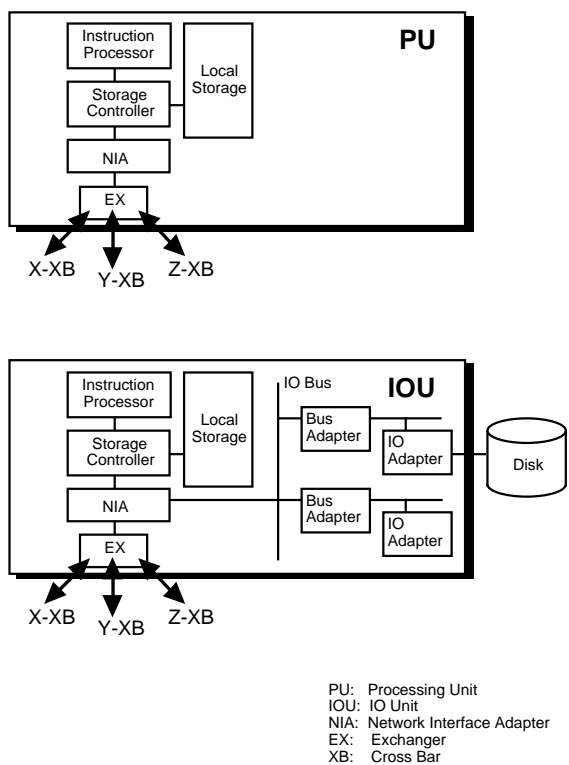}
\caption{Schematic configuration of Processing Unit(PU) board
and IO Unit(IOU) board of CP-PACS.}
\label{fig:BoardHard}
\end{figure}

On each processing unit(PU),
we place one enhanced PA-RISC processor, local
storage(DRAM) and a storage controller(see Fig.~\ref{fig:BoardHard}).
NIA stands for Network Interface 
Adapter and EX for exchanger. On an IO unit(IOU), in addition 
to the components on PU, we place an 
IO bus to which disks are connected through IO 
adapters.

\begin{figure}[t]
\epsfxsize=7.5cm \epsfbox{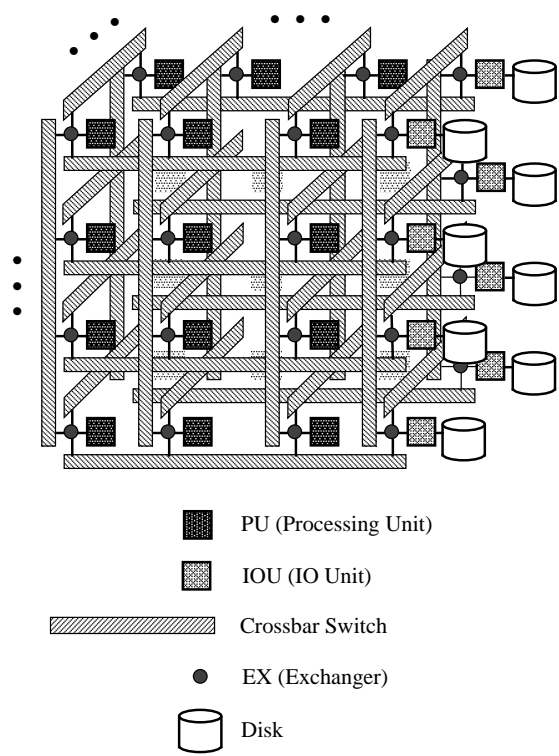}
\caption{System configuration of CP-PACS.}
\label{fig:Hxb}
\end{figure}
The network is
a 3-dimensional hyper crossbar as shown in Fig.~\ref{fig:Hxb}.
It consists of
x-direction crossbars as well as y and z direction crossbars. This hyper
crossbar network is very flexible: from any node to another node data
can be transferred through at most three switches.
The data transfer is made by message passing with wormhole routing.
The latency is expected to be of order of
a few micro-second.
A block-strided transfer is supported. 
We have also a global synchronization in addition to the hyper crossbar
network.

The system configuration of the CP-PACS with distributed disks is depicted in
Fig.~\ref{fig:Hxb}. 
The disk space is more than 500 Gbytes in total. We use RAID5
which has extra parity bits. 
In general, when the number of disks is large as in this case,
the MTBF(mean time between failure) becomes of order of one month.
With RAID there is no such problem, however.
The number of nodes, not fixed yet, is from 1000 to 1500.

The host is a main frame computer with modifications for massive
data transfer between the CP-PACS and the external disk storage.

A prototype with the PA-RISC without enhancement, which will be used
mainly for
tests of network hardware, will be completed in early 1994 and the
full scale machine with the newly developed processor is scheduled to be completed by
spring 1996.

The project is being carried out by a collaboration with Hitachi Ltd.
A new center called ``Center for Computational Physics''
was established at University of 
Tsukuba for the development of CP-PACS. 
A new building for the center, where 
the new machine will be installed,
was completed in the summer of 1993. The fund for the development of CP-PACS
is about \$14M.

\subsection{0.5-Teraflops project}
\begin{figure}[t]
\begin{center}
\leavevmode
\epsfysize=6.0cm
\epsfbox{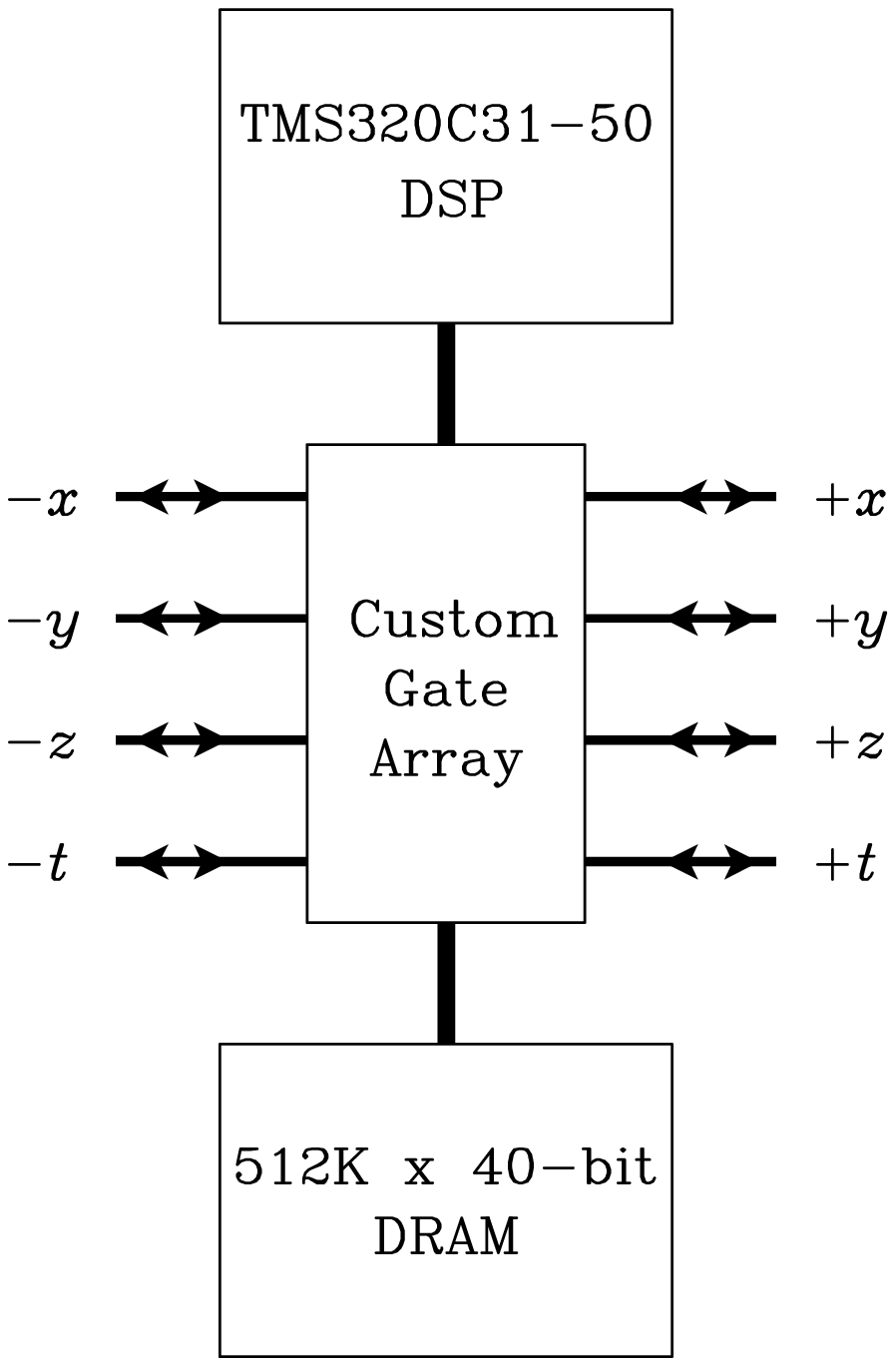}
\end{center}
\caption{Schematic diagram of one node of the 0.5-Teraflops machine.}
\label{fig:1norman}
\end{figure}
\begin{figure}[t]
\begin{center}
\leavevmode
\epsfxsize=5.0cm
\epsfbox{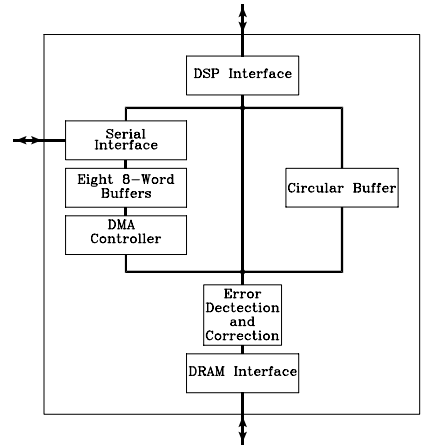}
\end{center}
\caption{Schematic diagram of NGA(node gate array) for the 0.5-Teraflops 
machine.}
\label{fig:3norman}
\end{figure}
\begin{figure}[t]
\begin{center}
\leavevmode
\epsfxsize=7.0cm
\epsfbox{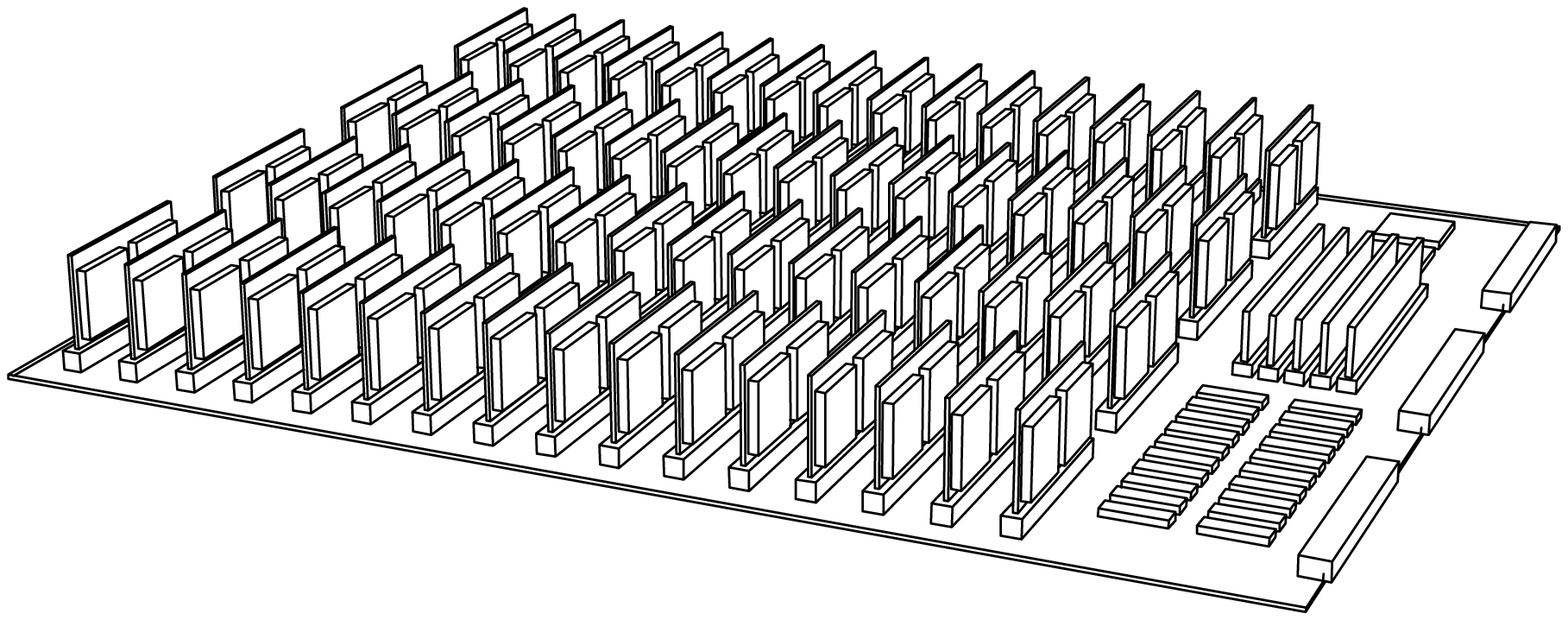}
\end{center}
\caption{Mechanical design of a mother board of the 0.5-Tflops project.}
\label{fig:2norman}
\end{figure}

This project started quite recently\cite{christ}.
The project is a collaboration of theoretical physicists and experimental
physicists\cite{0.5teraflops_mem}. The machine consists of 16K nodes making 
a 4-dimensional torus $16 \times 16 \times 16 \times 4$
with a peak speed of 0.8 Tflops with 32 bit arithmetic.
It is expected that the sustained speed for QCD is about 0.4 Tflops.

The node architecture is depicted in Fig.~\ref{fig:1norman}.
The processor is DSP(Digital Signal Processor) by Texas Instruments.
A 32 bit addition and multiplication can be 
performed concurrently with 40 ns machine cycle. This leads to 50 Mflops for
each node. 
It executes one word read for one machine cycle and one word write for two 
machine cycles. The DSP has 2K words of memory on chip. The size is 
small ($3.0$ cm$^2$),
the power consumption very low (less than 1 Watt) and
the price is less than 50\$.

Each node has 2 Mbytes of DRAM.
The maximum bandwidth between the processor and the memory is 25 Mwords/sec.
The memory size is 32 Gbytes in total.

The node gate array(NGA) which is shown in Fig.~\ref{fig:3norman}
is to be newly developed. The 
design 
has been partly finished. It plays the roles of memory
manager, network switch and specialized cache as a buffer. The buffer size
is chosen in such a way that multiplications of $3 \times 3$ matrices on 
3-vectors can be efficiently done.

The 4-dimensional network is connected by eight bi-directional lines of NGA.
Because the data transfer is made by handshaking, the latency is not low.
To hide this latency, there is a mode called ``store and pass through''.
In the calculation of the inner-product of two vectors which appears
in the conjugate gradient method, the data transfer which takes 70 \% of the total time
without this mode
reduces to 28 \% with this mode. It supports a block-strided transfer.

The mechanical design of a mother board is shown in Fig.~\ref{fig:2norman}. 
On the mother board there are $2 \times 2 \times 4 \times 4 = 64$ 
daughter boards
with last 4 making  a loop. Each node has 
a SCSI port to which peripheral tape and disk drives are connected.
One of 256 boards of the full machine is connected to the host.
The disk space is 48 Gbytes in total. The data transfer from disk to tape
or visa-visa can be done concurrently with physics calculations.

\begin{table*}[t]
\setlength{\tabcolsep}{0.5pc}
\caption{Characteristics of some commercial computers}
\label{tab:commercial}
\begin{tabular*}{\textwidth}{@{}l@{\extracolsep{\fill}}rrrr}
\hline
Machine & CM-5 & T3D & VPP500 & Paragon \\
\hline
processors & 1024 & 2048  & 222  & 4096 \\
\hline
arichi- & SIMD & MIMD & MIMD & MIMD \\
tecture  &+MIMD  &      &      &      \\
\hline
CPU         & SPARC & DEC & MCM &i860XP\\
            & +FPU  & Alpha& (custom) &      \\
\hline
Memory      & 32--128MB & 16(64)MB & 256MB & 32MB   \\
\hline
speed/ & & & &\\
processor & 128Mflops & 150Mflops & 1.6Gflops & 75Mflops\\
\hline
network & fat tree & 3d torus & crossbar & 2d mesh \\
\hline
host  & SUN WS & C90 & VP2600 & CONVEX\\
\hline
peak       &           &           &          &  \\
speed & 130 Gflops & 300 Gflops & 355 Gflops & 320 Gflops\\
\hline
data transfer   & 5-20 MB/sec& 300 MB/sec & 400 MB/sec & 200MB/sec\\
\hline
\end{tabular*}
\end{table*}

The power consumption is expected to be about 50 KW, which is very low compared
with other projects. The test board will be completed by summer 1994 and
the full machine by summer 1995. The funds for 128 node machine
with a peak speed of 6.4 Gflops is supported by DOE. The proposal for the full machine will be submitted in spring 1994.
\subsection{APE1000}
This is a successor of APE 100 with a peak speed of 1Tflops with 64 bit
arithmetic\cite{rapuano}. 
The project will start by the end of 1994.

\section{Commercial computers}
I list the characteristics of the most powerful commercial computers in
Table~\ref{tab:commercial} and describe in some details the two new ones below.
For other computers I refer to the earlier reviews\cite{review}.
\subsection{VPP500}
This is the latest machine from Fujitsu.
Each node is a vector processor with the same architecture as VP400
with a peak speed of 1.6 Gflops. 
Because of this, it is called a vector-parallel machine by Fujitsu.
One node is a multi-chip-module which consists of
121 LSIs, a part of which is composed of GaAs. 
Each node has 128 Kbytes of vector registers and 2 Kbytes of
mask registers. 
The memory size is 256Mbytes/node.
The network is a complete crossbar connecting all nodes,
which is very powerful for any
application.
The bandwidth for data transfer is 400 Mbytes/sec for each direction.
The OS is UNIX and the language is Fortran plus directives for parallel
procedures.

The maximum number of nodes is 222
with the peak speed of 355 Gflops.
The power consumption is 6KW/node. 
The power needed for the full machine is more than 1 MW.

A small VPP500 with 4 processors
is scheduled to be installed at Aachen this December.
Another one with 7 processors will be installed at the Institute of
Space and Astronomical Laboratory of Japan next January.
\subsection{T3D}
This is the machine just announced by CRAY.
The node processor is the DEC Alpha chip, 
which is one of the most powerful RISC chip in the market.
The clock cycle is 6.7ns and the peak speed of the chip is 150Mflops.
The memory size is 16Mbytes for one node with 4Mbit DRAM at present.
It will be upgraded soon to 64Mbytes with 16Mbit DRAM.
The memory is globally shared and physically distributed.

The network is a 3-dimensional torus. The bandwidth for data transfer
is 300MB/sec for each direction.
The latency of the communication is very low, less than 1 microsecond for hardware overhead.

It is a MIMD machine with a maximum peak speed of 300Gflops when it is composed
of 2048 nodes: the maximum number of nodes which is 1024 at present will be
increased to 2048 soon. 

The OS is Mach and the language is Cray Research Adaptive Fortran.

The machine with 32 nodes have been already installed at Pittsburgh
Supercomputing Center. It will be upgraded to 512 nodes next spring.

\begin{figure}[t]
\begin{center}
\leavevmode
\epsfxsize=7.0cm
\epsfbox{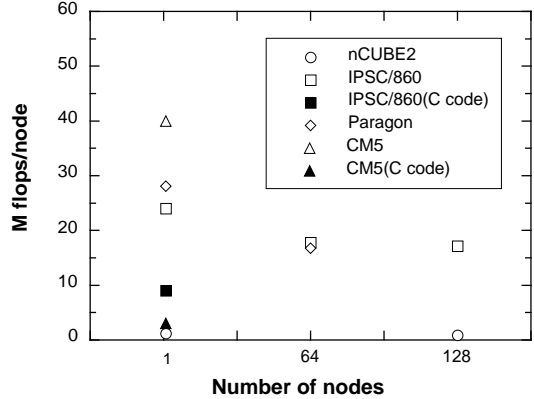}
\end{center}
\vspace{-0.5cm}
\caption{Sustained speed in terms of flops/node
of commercial parallel computers
for the conjugate gradient matrix inversion with staggered quarks[19].
The results for Paragon and CM5 are preliminary.}
\label{fig:commercial1}
\end{figure}

\begin{figure}[t]
\begin{center}
\leavevmode
\epsfxsize=7.0cm
\epsfbox{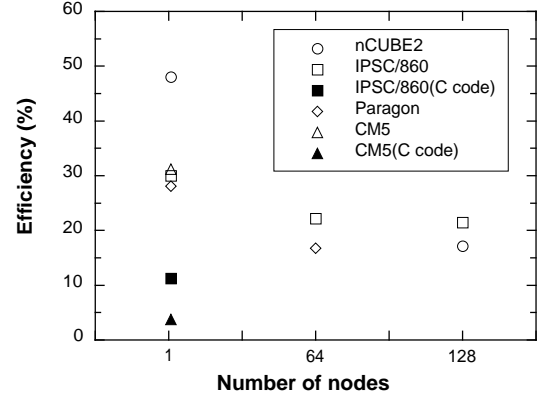}
\end{center}
\vspace{-0.5cm}
\caption{Efficiency in terms of the ratio of the sustained speed to the
theoretical peak speed[19].}
\label{fig:commercial2}
\end{figure}

\subsection{Sustained speed of commercial parallel computers}
The MILC collaboration has been running QCD codes
on a number of commercial computers including the nCUBE2, the Intel iPSC/860,
the Intel Paragon and the TMC CM5.
They have results of benchmarks 
for the conjugate gradient matrix inversion with staggered quarks
on these parallel computers\cite{sugar}.
The performances of the benchmarks are 
plotted in Figs.~\ref{fig:commercial1} 
and \ref{fig:commercial2}, respectively,
in terms of Mflops/node and the ratio of the sustained speed to the theoretical
peak speed. It should be noted that the benchmarks quoted for the CM5 and the
Paragon are preliminary. In particular, the communication speed of the
Paragon is expected to improve significantly as the operating system is 
upgraded.

The nCUBE2 is very stable and has nice software.
Because nCUBE2 is slow, it is not suitable for large QCD 
simulations, but it is convenient for software development.

When the code is written in C, 
the efficiency is very low for iPSC/860 and CM5
as is seen in the figures. Only when they
are written in assembly languages, the efficiency becomes around 30\%.
A similar efficiency has been also reported at this conference 
by Rajan Gupta\cite{gupta} for Wilson quarks in the case of CM5.

\section{Toward Teraflops computers}
\subsection{Three strategies}
Roughly speaking,
there are three strategies
to get a 1 Tflops machine
as shown in Table~\ref{tab:teraflops}.
\begin{table}[b]
\setlength{\tabcolsep}{0.5pc}
\caption{Towards 1 Teraflops machines}
\label{tab:teraflops}
\begin{tabular*}{7.4cm}{@{}l@{\extracolsep{\fill}}rr}
\hline

Speed of CPU & \#CPU&type\\
Mflops   & & \\
\hline
2000& 500&VPP500\\
 & & Teraflops\\
200--400&2,500--5,000&T3D, CP-PACS\\
50--100&10,000--20,000&0.5Teraflops,\\
 & &  CM5, nCUBE,\\
 & & Paragon\\
\hline
\end{tabular*}
\end{table}

The first is a vector-parallel approach taken by VPP500:
2 Gflops $\times$ 500 nodes =1 Tflops.
The second is the approach taken by T3D and CP-PACS, that is, to use
the most advanced RISC processor with an enhanced mechanism for high throughput
between memory and processor:
200-400 Mflops $\times$ 2500-5000 nodes = 1Tflops.
The approach taken by the Teraflops project is in between the first and the second
in the sense that the peak speed of one FPU is 200--300 Mflops and that of one
node is more than 1.6 Gflops.
The third approach is to use well-established technology taken by
CM5, Paragon, nCUBE and the 0.5-Tflops project:
50-100 Mflops $\times$ 10,000-20,000 nodes = 1 Tflops. 

In the first approach, the power consumption and the size will become 
problematical, although the number of nodes is small.
In the second approach, the sustained speed of each node
for arithmetic operations and that of the data
transfer between nodes will be the key issue.
In the third approach the packaging of the whole system and the reliability
will be crucial.
In spite of these potential obstacles, I believe that  the rapid progress
of technologies will enable all three approaches to reach 1 Tflops of
theoretical peak speed in a few years.  We should note, however, that achieving
a high sustained speed with massively parallel
computers and having flexibility for applications
require additional considerations on the balance of speed of various
components and other architectural issues.  Let us make brief comments on these
points.

\subsection{Balance of speed}

\begin{figure}[t]
\epsfxsize=7.5cm \epsfbox{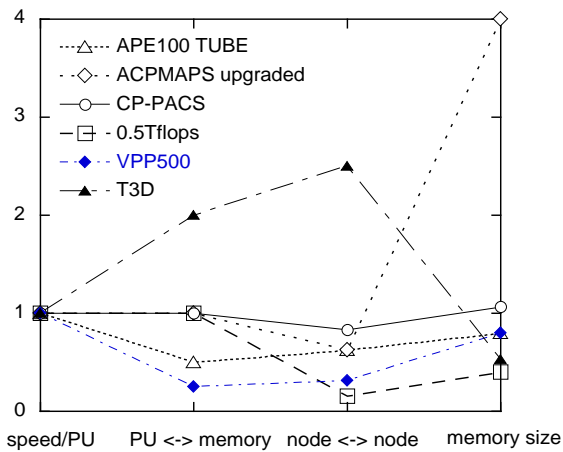}
\caption{Balance of bandwidth and memory size against processor speed.
The normalizations are
1 floating point operation/sec:0.5 words/sec:0.1 words/sec:0.025 words,
which is roughly the balance for lattice QCD.}
\label{fig:Balance}
\end{figure}

In Fig.~\ref{fig:Balance} the memory-processor bandwidth, the inter-node communication
bandwidth, and the memory size are compared against the processor speed for the
computers we reviewed in some detail.   The processor speed is normalized to
unity, and other normalizations are chosen for the following reason. 
For QCD calculation it is probably appropriate that
the bandwidth between CPU and the memory is one word for two floating point
operations.  
It also will be enough that the bandwidth for inter-node communication  is 
0.1 words for one floating point operation.
For the memory size, the 
normalization is arbitrary, and I chose 0.025 words of memory size for
1 flops/sec.

We see that each machine has its own characteristic.  Securing a
high bandwidth between memory and processor and that between nodes, sufficient to
keep up with the processor speed, is one of the crucial factor for a high
sustained speed.  In dedicated computer projects these parameters can be tuned to
specific applications (this in fact underlies the cost effectiveness of
dedicated computers).  For CP-PACS we have chosen the balance in such a way that
it is optimized for lattice QCD.  We should note, however, that the requirements
on the bandwidths in lattice QCD are modest compared to many other applications. 
Higher bandwidths are probably preferred for general purpose computers
as realized in the case of T3D.

There are other points which do not appear in the figure
such as the number of floating point registers on
each processor, the structure of memory (pipelined or not) and 
the latency of the communication.
These features are also important
for the performance of a massively parallel machine.
For example, the memory-processor bandwidth relative to the speed of one node
is small for VPP500, 
but it has 8Kbytes of registers which probably compensates it.

\subsection{Other issues of architecture}
\vspace{0.2cm}
\subsubsection{SIMD or MIMD}
SIMD is simple and generally sufficient for QCD calculations.  However, MIMD is
more flexible and can accommodate more varieties of algorithms.  An
interesting question is whether there are efficient algorithms for
inversion of quark matrices which requires a MIMD architecture.  Another point
is that  MIMD hardware is probably simpler than SIMD for a machine with a large
number of processors since the clock skew problem will become serious for SIMD.
\subsubsection{Topology of network}
The 3d torus and 4d torus networks are simple and natural for
lattice QCD. However, precision measurement of observables requires
finite-size analyses for which
we need simulations on a number of lattice sizes. For this point
more flexible network is preferable.
\subsubsection{32bit or 64bit}
In many cases of lattice QCD calculations it seems that 32bit arithmetic
is sufficient. However, for example, at the
global reject/accept step of the Hybrid Monte Carlo algorithm on a large lattice, 
the 32bit precision in not sufficient.
In general the 64 bit precision is needed when
the algorithm involves global variables.

\section{Conclusions}
In this review I have surveyed the development of parallel computers and the
present status of dedicated computer projects toward Teraflops of speed.  In the
1980's parallel computers were in their infancy and TMC was virtually the only
company in the field.  At that time there was no doubt that constructing
dedicated parallel computers by physicists was a beneficial project. In fact
dedicated computers which resulted from these
projects have produced a number of interesting and important physics results on
lattice field theories.  
The situation has become less clear-cut in recent years due to higher
technology needed to achieve faster speed on one part, and emergence
of powerful general purpose parallel computers from commercial vendors on the
other.

Historically projects for dedicated computers have been carried out by
a small group of lattice physicists, in some cases in collaboration with
experimental physicists and computer scientists, but without involvement of 
large commercial companies.   The 0.5-Teraflops project follows this spirit. 
Fully utilizing well-established  micro-processor technologies and designing aids
which have become commercially available,  the project aims to complete a
computer precisely tuned to lattice QCD within a short period of time and at a
low cost. It is very impressive to learn that this strategy is actually
possible for computers approaching a Teraflops of speed.  I believe that a vital
factor in realizing this approach is the experience gained with
the construction of three previous computers at Columbia. 

Another possible approach is to depart from the traditional style and to seek
for a close collaboration with large companies from the start of the project. 
This strategy is the one taken
by the US Teraflops project and the Japanese CP-PACS project.  In the
computers planned in these projects  the most advanced processors are to be
networked together with a large bandwidth.  The 0.5-micron semiconductor
technology, soon to become that of 0.3 micron, and the packaging technique
needed for this type of architecture can not be handled by 
physicists and computer scientists alone.  The cost is necessarily higher and
the  construction period longer.  There are, however, the advantage of choosing
more flexible architecture, reliability of hardware, and generally  better
software environment which is very important for development of application
programs and data analysis.  

Regardless of the approaches, I think dedicated computer projects still
represent an important avenue we should pursue for acquiring the computing power
needed for advancement of lattice field theory studies.  Hopefully all three
computers will be completed in a few years time and produce a variety of
fruitful results with some unexpected surprises.

\section*{Acknowledgments}
I am grateful to many colleagues for useful correspondence
and discussions.
I would particularly
like to thank N. Christ, M. Fischler, F. Karsch, R. Kenway, J. Negele,
F. Rapuano, R. Sugar A. Ukawa and D. Weingarten.
I am also indebted to the members of the CP-PACS
project, in particular K. Nakazawa.
Valuable suggestions of A. Ukawa on the manuscript are gratefully acknowledged.
Finally I would like to thank K. Kanaya, S. Aoki, H. Nakamura and H. Hirose
for the help in the
preparation of the manuscript. This work is supported in part by the
Grant-in-Aid of the Ministry of Education(No. 05NP0601).

\end{document}